\documentstyle[pra,aps,twocolumn,graphicx]{revtex}

\def\tpo#1{^{\otimes #1}}
\def\H{{\cal H}}
\def\B{{\cal B}}
\def\Cx{{\bf C}} 
\def\Rl{{\bf R}} 
\def\tr{{\rm tr}}
\def\idty{{\bf 1}}

\def\GLd{{\rm GL}(d,{\bf C})}

\def\SN{{\bf S}_N} 
\def\Cartan{C}
\def\diag{{\rm diag}}

\def\RR{{\cal R}} 
\def\SS{{\cal S}} 

\title{Estimating the spectrum of a density operator}
 \author{ M. Keyl\thanks{Electronic Mail: \tt{m.keyl@tu-bs.de}}
 {{}\ and\ } R.~F. Werner\thanks{Electronic Mail: \tt{r.werner@tu-bs.de}}
   \\[1ex]
  {\small Institut f{\"u}r Mathematische Physik, TU Braunschweig,}\\
  {\small Mendelssohnstr.3, 38106 Braunschweig, Germany.}}
\date{\today}

\begin{document}

\maketitle
\begin{abstract}
Given $N$ quantum systems prepared according to the same density
operator $\rho$, we propose a measurement on the $N$-fold system
which approximately yields the spectrum of $\rho$. The projections
of the proposed observable decompose the Hilbert space according
to the irreducible representations of the permutations on $N$
points, and are labeled by  Young frames, whose relative row
lengths estimate the eigenvalues of $\rho$ in decreasing order. We
show convergence of these estimates in the limit $N\to\infty$, and
that the probability for errors decreases exponentially with a
rate we compute explicitly.
\end{abstract}

\section{Introduction}

The density operator of a quantum system describes the preparation
of the system in all details relevant to statistical experiments.
Like a classical probability distribution it cannot be measured on
a single system, but can only be estimated on an ensemble sequence
of identically prepared systems. In fact, if we could determine
the density operator on a single quantum system, we could combine the
measurement with a device re-preparing several systems with the
measured density operator, in contradiction to the well-known
No-Cloning Theorem \cite{WooZu}. This points to a close connection
between the problem of estimating the density operator and approximate
cloning. In the case of inputs promised to be in a pure state the
optimal solutions to both problems are known \cite{Klo,Klo2,Bruss99},
and it turns out that in a sense the limit of the cloning problem for
output number $M\to\infty$ is equivalent to the estimation problem. The
``optimal'' cloning transformation was shown in this case to be
quite insensitive to the figure of merit defining optimality
\cite{Klo2}.

In the case of mixed input states much less is known about the
cloning problem. It is likely that in this case there may be
different natural figures of merit leading to inequivalent
``optimal'' solutions. Even the classical version the problem is
not trivial, and is related to the so-called bootstrap technique
\cite{Efron93} in classical statistics.

The estimation problem certainly has many solutions. In fact, any
procedure of determining the density matrix through the
measurement of the expectations of a suitable ``quorum'' of
observables \cite{Weigert00}, such as in quantum state tomography
\cite{Leonhardt97} is a solution. Other methods include adaptive
schemes \cite{Fischer00} where the result of one measurement is used to
select the next one. In all these cases, the estimate amounts to the
measurement of an observable on the full input state $\rho\tpo N$, which
factorizes into one-site observables. What we are concerned with here
is, as in the work of Vidal et. al. \cite{Vidal99}, the search for
improved estimates, admitting arbitrary observables on the $N$-fold
input  system, including ``entangled'' ones. In contrast to
\cite{Vidal99} however we are not interested in estimators which are
optimal for a more or less general figure of merit, but in the
asymtotic behaviour if the number $N$ of input systems goes to
infinity (in this context see also the work of Gill and Massar
\cite{GillMassar00}). 

When $\H\cong\Cx^d$ is the Hilbert space of a single system, the
overall input density operator of the estimation problem is
$\rho\tpo N$, which lives on the $N^{\rm th}$ tensor power $\H\tpo
N$. This space has a natural orthogonal decomposition according to
the irreducible representations of the permutation group of $N$
points, acting as the permutations of the tensor factors.
Equivalently, this is the decomposition according to the
irreducible representations of the unitary group on $\H$ (see
below). It is well-known that this orthogonal decomposition is
labeled by {\it Young frames}, i.e., by the arrangements of $N$
boxes into $d$ rows of lengths $Y_1\geq Y_2\geq\cdots,\geq
Y_d\geq0$ with $\sum_\alpha Y_\alpha=N$. There is a striking
similarity here with the spectra we want to estimate, which are
given by sequences of the eigenvalues of $\rho$, say, $r_1\geq
r_2,\cdots,\geq r_d\geq0$, with $\sum_\alpha r_\alpha=1$. The
basic idea of this paper is to show that this is not a superficial
similarity: measuring the Young frame (by an observable whose
eigenprojections are the projections in the orthogonal
decomposition) is, in fact, a good estimate of the spectrum. More
precisely, we show that the probability for the error $\vert
Y_\alpha/N-r_\alpha\vert$ to be larger than a fixed $\epsilon$ for
some $\alpha$ decreases exponentially as $N\to\infty$.

Our basic technique is the Theory of Large Deviations
\cite{Ellis85}, in particular a result by Duffield \cite{Duffield90}
on the Large Deviation properties of tensor powers of group
representations. This will allow us to compute the rate of
exponential convergence explicitly.

\section{Statement of the Result}
\label{sec:statement-result}

In order to state our result, explicitly, we need to recall the
decomposition theory for $N$-fold tensor products. Throughout, the
one-particle space $\H$ will be the $d$-dimensional Hilbert space
$\Cx^d$, with $d<\infty$. Two group representations play a crucial role:
firstly, the representation $X\mapsto X\tpo N$ of the general linear group
$\GLd$ and, secondly, the representation $p\mapsto S_p$ of the permutations
$p\in\SN$ on $N$ points, represented by permuting the tensor factors: 
\begin{equation}\label{permute}
 S_p\ \psi_{1}\otimes\cdots \psi_{N}
  =\psi_{p^{-1}1}\otimes\cdots \psi_{p^{-1}N}\;.
\end{equation}
The basic result \cite{Simon96} is that these two representations
are ``commutants'' of each other, i.e., any operator on $\H\tpo N$
commuting with all $X\tpo N$ is a linear combination of the $S_p$,
and conversely. This leads to the decomposition
\begin{eqnarray}\label{decomp}
   \H\tpo N&\cong& \bigoplus_Y \RR_Y\otimes\SS_Y \\
    X\tpo N&\cong& \bigoplus_Y \pi_Y(X)\otimes \idty \label{eq:2}   \\
    S_p    &\cong& \bigoplus_Y \idty\otimes \widehat\pi_Y(p) \label{eq:3} \;,
\end{eqnarray}
where $\pi_Y:\GLd\to\B(\RR_Y)$ and $\widehat\pi_Y:\SN\to\B(\SS_Y)$
are irreducible representations, and the restriction of $\pi_Y$ to
unitary operators is unitary. The summation index $Y$ runs over
all Young frames with $d$ rows and $N$ boxes, as described in the
introduction. We denote by $P_Y$ the projection onto the
corresponding summand in the above decomposition.

Let us consider now the estimation problem. As already discussed
in the introduction, we are searching for an observable $E_N$
describing a measurement on $N$ $d$-level systems, whose readouts
are possible spectra of $d$-level density operators. The set of
possible spectra will be denoted by
\begin{equation}
  \Sigma = \{ s \in \Rl^d \, | \, x \triangleright 0, \ \mbox{$\sum_{j=1}^d x_j$} = 1\}
\end{equation}
where $x \triangleright 0$ denotes the ordering relation on
$\Rl^d$ given by
\begin{equation} \label{eq:9}
  s \triangleright 0 :\iff s_j > s_{j+1} \ \mbox{for all} \ j=1,\ldots,d-1.
\end{equation}
Technically, $E_N$ must be a positive operator valued measure on
this set, assigning to each measurable subset
$\Delta\subset\Sigma$ a positive operator
$E_N(\Delta)\in\B(\H^{\otimes N})$, whose expectation in any given
state is interpreted as the probability for the measurement to
yield a result $s\in\Delta$.

The citerion for a good estimator $E_N$ is that, for any
one-particle density operator $\rho$, the value measured on a
state $\rho\tpo N$ is likely to be close to the true spectrum
$r\in\Sigma$ of $\rho$, i.e., that the probability
\begin{equation}\label{eq:1}
  K_N(\Delta) := \tr\bigl(E_N(\Delta)\rho^{\otimes N}\bigr)
\end{equation}
is small when $\Delta$ is the complement of a small ball around
$r$. Of course, we will look at this problem for large $N$. So our
task is to find a whole sequence of observables $E_N$,
$N=1,2,\ldots$, making error probabilities like (\ref{eq:1})
go to zero as $N\to\infty$.

The search for efficient estimation strategies $E_N$ can be
simplified greatly by symmetry arguments. To see this, consider a
permutation $p \in \SN$. If we insert the transformed estimator
$S_p E_N(\Delta) S_p^*$ into Equation (\ref{eq:1}) we see
immediately that $K_N(\Delta)$ remains unchanged. Replacing
$E_N(\Delta)$ by the average $N!^{-1} \sum_{p \in \SN} S_p
E_N(\Delta) S_p^*$ shows that we may assume $[E_N(\Delta), S_p] =
0$ for all permutations $p$, without loss of estimation quality. A
similar argument together with the fact that the quality of the
estimate is judged by some criterion not depending on the choice
of a basis in $\H$ shows that we may assume in addition that
$E_N(\Delta)$ commutes with all unitaries $U^{\otimes N}$. But
this implies according to Eq.~(\ref{eq:2}) and (\ref{eq:3}) that
$E_N(\Delta)$ must be a function of the projection operators
 $P_Y: \H\tpo N \to \RR_Y \otimes \SS_Y$
defined at the beginning of this section. If we require in
addition that each $E_N(\Delta)$ be a projection, which is
suggestive for ruling out unnecessary fuzziness, $E_N$ must be of
the form
\begin{equation} \label{eq:4}
  E_N(\Delta) = \sum_{Y:s_N(Y)\in\Delta}  P_Y ,
\end{equation}
where $s_N$ is an arbitrary mapping assigning to each Young frame
$Y$ (with $d$ rows and $N$ boxes) an estimate $s_N(Y)\in\Sigma$.
In other words, the estimation proceeds by first measuring the
Young frame projections $P_Y$ and then computing an estimate
$s_N(Y)$ on the basis of the result $Y$.

\begin{figure}[t]
  \begin{center}
    \includegraphics[scale=0.75]{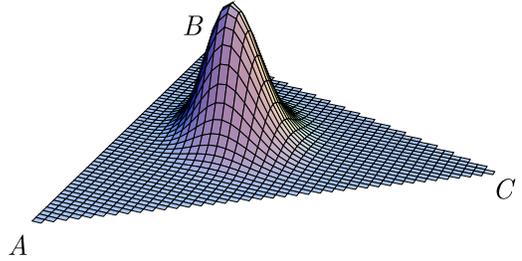}
    \caption{Probability distribution $\tr(\rho\tpo NP_Y)$ for $d=3$,
      $N=120$ and $r=(0.6, 0.3, 0.1)$. The set $\Sigma$ is the triangle
      with corners $A=(1,0,0)$, $B=(1/2,1/2,0)$, $C=(1/3, 1/3,
      1/3)$.} 
    \label{Fig1}
  \end{center}
\end{figure}

The simplest choice is clearly to take the normalized Young frames
themselves as the estimate, i.e.,
\begin{equation} \label{eq:6}
  s_N(Y) = Y/N.
\end{equation}
It turns out somewhat surprisingly that with this choice the
$E_N(\Delta)$ from Eq.~(\ref{eq:4}) form an asymptotically exact
estimator. By this we mean that, for every $\rho$, the probability
measures $K_N$ from Eq.~(\ref{eq:1}) converge weakly to the point
measure at the spectrum $r$ of $\rho$. Explicitly, for each
continuous function $f$ on $\Sigma$ we have
 \begin{eqnarray}
   \lim_{N \to \infty} &\int_\Sigma& f(s)K_N(ds) =
   \nonumber\\
   &=& \lim_{N\to\infty} \sum_Y f\left(\frac{Y}{N}\right) \tr\bigl(\rho^{\otimes N}
   P_Y\bigr)  = f(r) \label{eq:7}.
 \end{eqnarray}

We illustrate this in Figure~\ref{Fig1}, for $d=3$, $N=120$, and $\rho$ a
density operator with spectrum $r=(0.6, 0.3, 0.1)$. Then $\Sigma$ is a
triangle with corners $A=(1,0,0)$, $B=(1/2,1/2,0)$, and
$C=(1/3,1/3,1/3)$, and we plot the probabilities $\tr(\rho\tpo NP_Y)$
over $Y/N\in\Sigma$. The explicit computation uses the Weyl character
formula~\cite[IX.9.1]{Simon96}, which we do not need elsewhere in the
paper.  

Clearly, the distribution is peaked at the true spectrum and our claim
is that this will become exact in the limit $N\to\infty$. To prove
convergence we will use large deviation methods which give us not only
the convergence just stated but an {\em exponential error estimate} of
the form 
\begin{equation}\label{KNexpo}
  K_N(\Delta) \approx \exp\left(- N \inf_{s \in \Delta} I(s)\right),
\end{equation}
where $I$ denotes a positive function on $\Sigma$, called the
\emph{rate function}, which vanishes only for $s=r$.

For the statement of the main theorem we say that a measurable
subset $\Delta\subset\Sigma$ has ``small boundary'', if its
interior is dense in its closure. A typical choice for $\Delta$ is
the complement of a ball around the true spectrum.

\proclaim Theorem. The estimator defined in Eqs. (\ref{eq:4}) and
(\ref{eq:6}) is asymptotically exact. Moreover, we have the error
estimate
\begin{equation}
  \lim_{N\to\infty} \frac{1}{N} \ln\;K_N(\Delta)
    = \inf_{s \in \Delta} I(s),
\end{equation}
for any set $\Delta\subset\Sigma$ with small boundary, where the
rate function $I: \Sigma \to [0,\infty]$ is
\begin{equation} \label{eq:8}
  I(s) = \sum_j s_j\left(\ln s_j - \ln r_j\right).
\end{equation}

The expression for $I$ is the relative entropy \cite{OhyaPetz} of the 
probability vectors $s$ and $r$. Relative entropies occur also as
the rate functions in Large Deviation properties of independent
identically distributed (classical \cite{Cramer38} or quantum
\cite{Werner92}) random variables, although there seems to be no
direct way to reduce the above Theorem to these standard setups.

\section{Sketch of Proof}

Rather than giving a  proof of every detail, our aim here is to
explain why the the scaled Young frames $Y/N$ appear in the
estimation problem. The crucial observation is that the Young
frame $(Y_1,\ldots,Y_d)$ is the {\it highest weight} of the
representation $\pi_Y$ in the ordering $\triangleright$ and this ordering is directly
related to picking out the fastest growing exponential in certain
integrals of the measures $K_N$. 

The integrals we need to study are the Laplace transforms of the
measures $K_N$. We introduce the ``scaled cumulant
generating function''
\begin{equation}\label{cmu}
  c(\eta)=\lim_{N\to\infty}\frac{1}{N}\ \ln\
         \int_\Sigma K_N(ds)e^{N\eta\cdot s}\;,
\end{equation}
where $\eta\in\Rl^d$, and $\eta\cdot s$ is the scalar product. If
the measures $K_N$ behave like Eq.~(\ref{KNexpo}) the integrand near
$s$ behaves like $\exp N(\eta\cdot s-I(s))$, and the largest contribution
comes from the fastest growing exponential: 
\begin{equation}\label{Varadhan}
  c(\eta)=\sup_s(\eta\cdot s-I(s))\;.
\end{equation}
This is an instance of Varadhan's Theorem \cite{Varadhan66}, which has a
converse, the G{\"a}rtner-Ellis Theorem \cite[Thm. II.6.1]{Ellis85}: if
the limit (\ref{cmu}) exists, and is differentiable then the estimate
in the Theorem holds, with the rate function determined from
(\ref{Varadhan}) by inverse Legendre transformation. We will follow
Duffield \cite{Duffield90} by computing the limit (\ref{cmu}) from
group theoretical data. 

Consider the ``maximally abelian subgroup'' $\Cartan\subset\GLd$
of diagonal matrices
\begin{equation} \label{eq:14}
  \rho_h =\diag(\exp(h_1),\ldots,\exp(h_d))
\end{equation}
for $h\in\Cx^d$. Since these commute, all the operators
$\pi_Y(\rho_h)$ commute in every representation $\pi_Y$, and can
hence be simultaneously diagonalized. The vectors
$\mu=(\mu_1,\ldots,\mu_d)$ such that $\pi_Y(\rho_h)\psi=\exp(\mu
\cdot h)\psi$ for some non-zero vector $\psi$ are called {\it weights} of
the representation $\pi_Y$. The dimension $m(\mu)$ of the
corresponding eigenspace is called the {\em multiplicity} of
$\mu$. One particular weight (with multiplicity one) is the Young
frame $Y$ itself (interpreted as an element of $\Rl^d$) and it
turns out that $Y$ is the maximum (the ``heighest weight'') among
all weights of $\pi_Y$, in the $\triangleright$-ordering from Equation
(\ref{eq:9}). Representation theory of semisimple Lie algebras
\cite{Simon96} shows that each irreducible, analytic representation of
$\GLd$ is uniquely characterized (up to unitary equivalence) by its
highest weight $Y$. 

In order to estimate the integral (\ref{cmu}), we need the
quantities $\tr(\rho^{\otimes N} P_Y)$. For simplicity we assume
that $\rho$ is non-singular, i.e.,  an element of
$\GLd$. By Equation (\ref{eq:2}) we have
\begin{equation}
  \tr(\rho\tpo N P_Y) = \tr\bigl(\pi_Y(\rho) \otimes \idty \bigl)
  = \chi_Y(\rho) \dim(\SS_Y),
\end{equation}
where
\begin{equation}
  \chi_Y(\rho) := \tr\bigl(\pi_Y(\rho)\bigr)
\end{equation}
is the {\em character} of the representation $\pi_Y$. Since
$\chi_Y$ is unitarily invariant
 ($\chi_Y(U\rho U^*) = \chi_Y(\rho)$) we may assume without loss of generality that
$\rho$ is diagonal and its matrix elements are arranged in
descending order. Using the notation from Equation (\ref{eq:14})
this assumption reads
\begin{equation} \label{eq:13}
  \rho = \rho_h \in \Cartan \ \mbox{with} \ h \triangleright 0 \
            \mbox{and} \ \sum_j \exp(h_j)
  = 1.
\end{equation}
Hence we can express $\chi_Y(\rho)$ in terms of the weights of $\pi_Y$:
\begin{equation} \label{eq:11}
  \chi_Y(\rho) = \sum_\mu m(\mu) \exp(\mu \cdot h),
\end{equation}
where the sum is taken over all weights $\mu$ of $\pi_Y$. Since $h
\triangleright 0$ and $Y\triangleright\mu$ for all $\mu$  we see
that $\exp(Y \cdot h)$ is the largest exponential. We therefore
estimate
\begin{equation} \label{eq:15}
  \exp(Y \cdot h) \leq \chi_Y(\rho)
    \leq \dim\left(\RR_Y\right) \exp(Y \cdot h).
\end{equation}
We will combine this with the consequence of Weyl's dimension formula
that $\dim\left(\RR_Y\right)$ is bounded above by a polynomial $p(N)$
in $N$, uniformly in $Y$  \cite[Lemma 2.2]{Duffield90}.
Hence, for any $h,\eta\in\Rl^d, h,\eta\triangleright0$ the two
expressions
\begin{eqnarray}\label{J}
  J(h,\eta)&=&\int_\Sigma K_N(ds)e^{N\eta\cdot s}
   =\sum_Y\tr(\rho_h\tpo NP_Y)e^{N\eta\cdot Y/N}\nonumber\\
   &=&\sum_Y\chi_Y(\rho_h)\;e^{\eta\cdot Y}\dim(\SS_Y)
\\ \noalign{and}\nonumber\\
  J'(h,\eta)&=&\sum_Ye^{(h+\eta)\cdot Y}\dim(\SS_Y)
\end{eqnarray}
are asymptotically equivalent in the sense that
 $(1/N)(\ln J(h,\eta)-\ln J'(h,\eta))\to0$. In the same sense we
can continue the chain of equivalences
\begin{eqnarray}
 J(h,\eta)&\approx& J'(h,\eta)=J'(h+\eta,0)
          \approx J(h+\eta,0)=\nonumber\\
          &=&\tr(\rho_{h+\eta}\tpo N)=(\tr\rho_{h+\eta})^N\;.
\end{eqnarray}
Hence, if $r_\alpha=\exp({h_\alpha})$ are the eigenvalues of a
non-singular density operator, we get for Eq.~(\ref{cmu}) the
expression
\begin{equation}\label{ccomputed}
  c(\eta)=\ln \sum_\alpha r_\alpha \exp(\eta_\alpha)
\end{equation}
It is then a simple calculus exercise to verify the above rate
function as the Legendre transform $I(s)=\sup_\eta(\eta\cdot
s-c(\eta))$.

This concludes our sketch of proof. In order to expand it into a
full proof, one needs to extend the computation of $c(\eta)$ to
$\eta/\mkern-13mu\triangleright0$, and prove that this extension
has the required regularity properties for the application of the
G{\"a}rtner-Ellis Theorem cited above. This has been carried
out by Duffield \cite{Duffield90} in a context which is on the one
hand wider, because it includes tensor powers of much more general
representations of semisimple Lie groups, but on the other hand is
narrower, because it contains only the case $\rho=d^{-1}\idty$ of
our Theorem. However, one can extend Duffield's result by
multiplying his measures $K_N$ with the factor $\chi_Y(\rho)/\chi_Y(\idty)$,
and using for this factor the estimate~(\ref{eq:15}).

\section{Discussion}

Although the estimate we discuss is asymptotically exact, it is
not at all clear whether and in what sense it might be {\it
optimal}, even for finite $N$. We have experimented with various
figures of merit for estimation and found different ``optimal''
estimators for low $N$, rarely coinciding with the $E_N$
determined by (\ref{eq:6}). It is also not at all clear how much
could be gained by optimization here.

An interesting extension will also be the construction of
estimators for the full density operator. It is very suggestive to
compose this out of the above estimator for the spectrum, and to
use for each Young frame a covariant observable to estimate the
eigenbasis of $\rho$. The density of the covariant observable
might be based on the highest weight vector of $\pi_Y$.

\section*{Acknowledgements}

Funding by the European Union project EQUIP (contract IST-1999-11053)
and financial support from the DFG (Bonn) is greatfully acknowledged.


\end{document}